\documentclass[letterpaper,twoside,preprintnumbers,slac_one]{revtex4}
\usepackage{graphicx}
\usepackage{fancyhdr}
\usepackage{amsmath} % American Mathematics Society standards
\usepackage{bm}% bold math
\usepackage{amsxtra}
\usepackage{amssymb}
\usepackage{amsthm}
\usepackage{latexsym}
\usepackage{lscape}

\renewcommand{\headrulewidth}{0pt}
\renewcommand{\footrulewidth}{0pt}

\usepackage{calc}   % working with lengths, counters etc.
\usepackage[        %
  includeheadfoot,  %
  textheight=235mm, %
  textwidth=170mm   %
]{geometry}         % set page layout parameters

\usepackage{xspace}
\usepackage[colorlinks=true,urlcolor=blue,linktocpage=true]{hyperref}
\usepackage{hyperref}
\hypersetup{
  pdftitle    = {Spontaneous Symmetry Breaking},
  pdfauthor   = {Gerald S. Guralnik <gerry@het.brown.edu>},
  pdfsubject  = {High Energy Physics - Theory},
  pdfcreator  = {LaTeX2e with hyperref package},
  pdfproducer = {pdflatex},
  pdfkeywords = {Quantum Field Theory, Schwinger-Dyson Equations, Lattice,
    Symmetry Breaking, Non-Perturbative QFT},
  pdfview     = {FitH}
}
\usepackage{ucs}
\usepackage[utf8x]{inputenc}
\usepackage[T1]{fontenc}
\usepackage{ae,aecompl}
\usepackage{lmodern}
\usepackage{microtype}
\usepackage[x11names]{xcolor}
\usepackage[american]{babel}
\def\hksqrt{\mathpalette\DHLhksqrt}
\def\DHLhksqrt#1#2{\setbox0=\hbox{$#1\sqrt{#2\,}$}\dimen0=\ht0
  \advance\dimen0-0.2\ht0
  \setbox2=\hbox{\vrule height\ht0 depth -\dimen0}%
{\box0\lower0.4pt\box2}}
\makeatletter
\def\equalsfill{$\m@th\mathord=\mkern-7mu
  \cleaders\hbox{$\!\mathord=\!$}\hfill
  \mkern-7mu\mathord=$}
\makeatother
\newcommand{\pa}{\partial}
\newcommand{\comm}[2]{\left[\,#1,#2\,\right]}

\newcommand{\ipop}[3]{\ensuremath{\langle#1 | #2 | #3\rangle}\xspace}

\newcommand{\bo}{\raise+0.0mm\hbox{$\Box$}}
\begin{document}

% Institutional report number...
\preprint{BROWN-HET-1618}

\title{The Beginnings of Spontaneous Symmetry Breaking in Particle
  Physics --- Derived From My on the Spot ``Intellectual Battlefield Impressions''}

\author{G.~S. Guralnik}
\affiliation{Department of Physics, Brown University, Providence, RI, USA}
\email{gerry@het.brown.edu}
\homepage{http://www.het.brown.edu/}
\date{2011-Sep-30}
\begin{abstract}
I summarize the development of the ideas of Spontaneous symmetry
breaking in the 1960s with an outline of the Guralnik, Hagen, Kibble
(GHK) paper and include comments on the relationship of this paper to
those of Brout, Englert and Higgs. I include some pictures from my ``physics
family album which might be of some amusement value''
\smallskip

Institutional Report Number: BROWN-HET-1618.
\end{abstract}

\maketitle

\thispagestyle{fancy}


%%%%%%%%%%%%%%%%%%%%%%%%%%%%%%%%%%
\section{Introduction}\label{sec:0}
This paper borrows freely from my review in IJMPA \cite{gg;2009} as
well as previous talks given by Hagen, Kibble and myself. I give an
overview of the status of particle physics in the 1960s with
particular focus on the work that I did with Richard Hagen and Tom
Kibble (GHK) \cite{ghk;1964}. Our group and two others, Englert-Brout
(EB) \cite{eb;1964} and Higgs (H) \cite{phpl;1964,ph;1964} worked on
what is now commonly refered to as the ``Higgs'' phenomenon and/or a
specific example of this constructed through the spontaneously broken
scalar electrodynamics. This work eventually led to the unified theory
of electroweak interaction as developed by Weinberg and Salam
\cite{sw;1967,as;1967}. I discuss how I understand symmetry
breaking. My overall viewpoint has changed little in basics over
the years. However, with the hope of clarifying this viewpoint, I mention some
more modern ideas which involve the understanding of the full range of
solutions of quantum field theories \cite{ggg;1996,ggzg;2007}. Later
in this document, in response to statements that have been made in
print and at other conferences, I discuss the differences between our
work and that of EB and H.
\section{What was theoretical particle physics like in the early 1960s?} \label{sec:1}
It is ironic that just before the `` Golden age'' of particle physics
began to unfold in the 1960s, Quantum Field theory was widely thought
to have reached a dead end. There was no understanding as to how to go
beyond coupling constant perturbation and thus there seemed to be
little hope of performing detailed calculations of strong interactions. As a result but
for a few places, including Harvard, MIT and Imperial college (the GHK
home institutions), most theoretical research attention was focussed
on S Matrix theory.

Fortunately, with the notable exception of lattice computational
methods, the basic field theory tools and formulations we use today
were, for the most part, available in some form. The Schwinger Action
Principle with associated Green's Function methods were well
developed, but mostly used by physicists who were in Schwinger's
Harvard centered following.  The (equivalent) Feynman Path Integral
path integral method also existed although its application was
probably even less known than Schwinger's methods. That either of
these methods could easily lead to solutions of quantum field theory
that were fundamentally non (coupling constant) perturbative was, at
the beginning of the 60s, at most an idea in waiting. This idea was
soon to be developed with the introduction of the concept of
spontaneous symmetry breaking. Progress still continues as illustrated
by the fact that only recently was it understood that path integrals
must include paths extended into the complex plane for them to
encompass the extended range of solutions to quantum field theory
\cite{ggg;1996,ggzg;2007}. Coupling constant perturbation theory was
well understood and widely applied as the major computation tool of
quantum field theory.  As an essential adjunct, renormalization theory
was well developed and had enjoyed sensational success in application
to problems involving fermions interacting
electromagnetically. However, it is of relevance to note that many
physicists were very unclear as to how to work with operators in
quantum electrodynamics as was evidenced by a fairly constant stream
of papers using cumbersome indefinite metric formulations. This was
not an issue if Schwinger's methods of quantizing in the radiation
gauge were followed. In the end, of course, everyone ended up
calculating the same graphs but on the way to this there were many
ways to make errors of interpretations.

There was deep concern about the four-fermion interaction not being
renormalizable since this cast serious doubt on our understanding of
weak interactions. This was particularly frustrating since the lowest
order perturbative calculations using four fermion interactions
described many weak processes with surprising accuracy. There was some
experimentation with summation techniques of subsets of graphs with
hope of understanding this problem.

Very importantly, we had the beginnings of the Standard Model.
The idea of electromagnetic like interactions of multiplets of fields
was introduced with the ideas of Yang, Mills and Salam's student Shaw
\cite{ym;1954,ss;1955} in 1954.  J.J Sakurai \cite{sak;1960} had
suggested the fundamental importance of vector interactions to a basic
model of particle interaction.

Essential to putting this all together was the contribution of the V-A theory of the weak
interactions by Sudarshan and Marshak (1957) which was according to Feynman in 1963 ``publicized by
Feynman and Gell-Mann'' \cite{gmf;1958}. The realization that the V-A combination is
the valid way to describe the weak interactions is essential to the
unified Electroweak theory and the standard model.

The importance of group theory to categorizing the elementary
particles was approximate flavor $SU(3)$ was well appreciated due to
the work of Murray Gell-Mann \cite{mg;1962} and Y. Ne'mann
\cite{yn;1961} (another Salam Student) and regarded as largely
confirmed thanks to the discovery in 1963 of the $\Omega^{-}$ which
was needed to fill in the baryon decuplet (10 particles).

\begin{center}
  \includegraphics[scale=0.5]{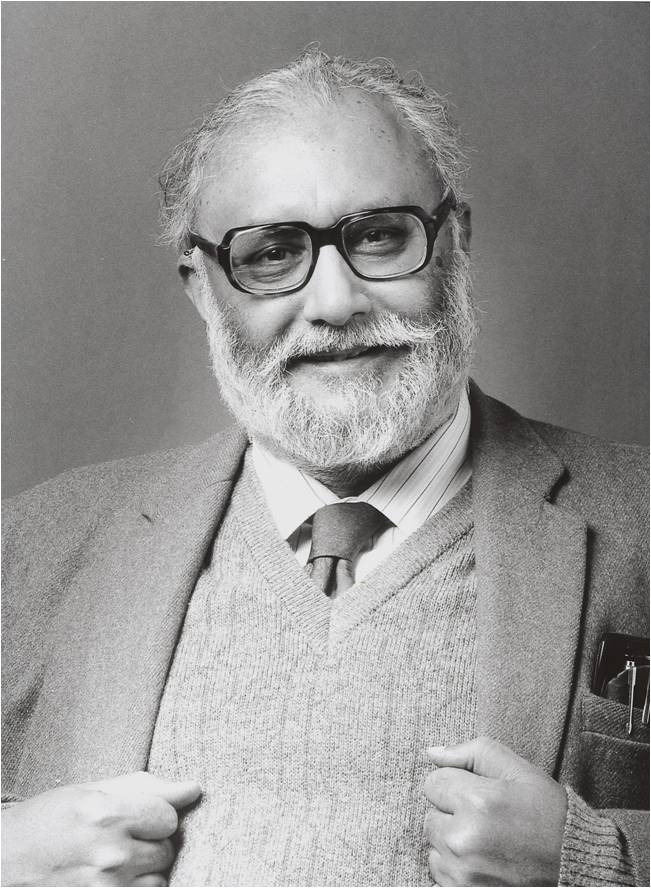}

  Abdus Salam
\end{center}

The Gell-Mann--Zweig quark (aces) ideas were formulated by 1964
\cite{gm;1964,gz;1964} but were far from completely accepted. The idea
of fundamental quarks became theoretically feasible due to the 1963
work by Wally Greenberg \cite{wag;1964}. His ideas lead to the concept
of color and made provided a workable way to construct fermions out of
3 quarks.

The story I want to focus on began in 1961 when Nambu with
Jona-Lasinio (NJL)  published \cite{nb;1960,nb;1961} their study of
spontaneous symmetry breaking of chiral symmetry using the four fermion
interaction
\begin{equation*}
  g\, \left[ (\bar{\psi}\psi)^2 - (\bar{\psi}\gamma_{5}\psi)^2 \right].
\end{equation*}
This interaction in itself was disturbing (just as was the four
fermion interaction to describe weak processes) because perturbation
theory with expansion around $g=0$ produces a series of increasingly
divergent terms that cannot be renormalized. NJL studied this model by
imposing a constraint that, at first glance, seemed to be inconsistent
with its symmetry and then formulated a new (not coupling constant
perturbation theory) leading order approximation. This formulation led
to a constraint that allowed consistency if a zero mass pseudoscalar
boson appeared in the calculated particle spectrum. This is the first
example of the Nambu- Goldstone boson.  NJL argued, without an
entirely convincing proof, that the massless particle is an exact
requirement following from the imposition of dynamical symmetry
breaking. This was a truly major breakthrough with new physics and a
new way at looking at solutions of relativistic quantum field theory!
We understood little beyond coupling constant perturbation theory
about how to actually solve a QFT. Attempts had been made to re-sum
these series but until NJL there was little or no realization that one
should be looking for an entirely different solutions and that these solutions could violate
obvious symmetries of the action.

After the NJL papers, J. Goldstone wrote (1961) \cite{jg;1961} his
famous paper where he examined a two (real) component scalar field
theory with a quartic self interaction. This presented a much simpler
illustration of a spontaneous symmetry breaking than that of NJL.
Here the interaction has a conserved charge symmetry which is
dynamically broken by requiring that the vacuum expectation of the
scalar fields non-zero. In this case, the leading
order approximation to the broken symmetric solution can be understood
without dealing with constrained divergent integrals. A constraint, of
course, is still present but it occurs as a simple numerical relation
(to leading order). While the original Lagrangian contains two equal
mass scalar fields, the resulting solutions (again and importantly to
leading order) look like those of a free field with two scalar bosons
of different mass with one mass vanishing and the square magnitude of
the other mass turning out to be twice the magnitude of the original
``bare mass''. Goldstone speculates that the zero mass excitation is required to all orders.
In fact,  higher order calculations result in graphical
structures based on the leading order propagators and the massless
particle stay massless while the mass of the other particle must be
remormalized and, of course, varies order by order.

In 1962 Goldstone Salam and Weinberg \cite{gsw;1962} convincingly proved that the
spontaneous breaking of a continuous global symmetry in a relativistic
theory requires associated zero mass excitations. This was a disappointment as they had hoped to
find that there was a way around this connection. Zero mass particles in nature are simply not seen with the
exception of the photon and presumably the graviton. Consequently, the idea of spontaneous symmetry breaking
seemed like it could not lead to new physics. Describing how this apparent problem was solved is the main point of
this paper. But before I get to this I will indicate why the whole
appearance of addition solutions to Quantum Field theory (QFT) beyond those
given by perturbation theory are to be expected and are in fact
suggested (and explained) by the solution set of ordinary differential differential equations. This
analysis is relatively new, but I think helps to clarify what is happening here even for those without
much detail knowledge of the mathematics of QFT.

The beginning point to understand the multiplicity of solutions to QFT
is the 1952 work of Dyson \cite{dyson;1952}. This paper showed that
there is a singularity at zero coupling in QED, thus giving an early
indicator that perturbation theory could not be the whole story and
indeed that the perturbation expansion is asymptotic not
convergent. As mentioned above, we can get insight by examining simple
differential equations. As a starter, it is not hard to see, using the Feynman path
integral or the Schwinger action principle, that such equations can be
interpreted as describing a zero dimensional QFT. As an explicit example, consider,

\begin{equation*}
  g\, \frac{d^3 y}{d J^3} + m^2\, \frac{d y}{d J} = J \; .
\end{equation*}

How many solutions does this have? Three! They can be found from an integral representation (zero
dimensional Feynman path integral for quartic interacting scalar field
theory):
\begin{equation*}
  Z = \int\, e^{-g\, \frac{\phi^4}{4} - \frac{m^2}{2}\,\phi^{2} + J\, \phi}\, \mathcal{D}\phi
\end{equation*}

The $J$ derivative of $Z$ satisfies the original differential equation
when the integral is evaluated over COMPLEX paths which do not
contribute at the end points. It is easy to show that there are 3 allowed independent paths in the
complex plane. Associated with the three integration paths, the integral has 3
stationary points that correspond to the three solutions of the
original differential equation.

These are $(J = 0)$ located at
\begin{equation*}
  \phi = 0, \; \phi = \pm\hksqrt{\frac{-m^2}{g}} \; .
\end{equation*}

Expanding around these stationary phase points to discover
asymptotic expressions for each of the three solutions is straightforwar. The path along
the real axis corresponds to the stationary point at $\phi=0$ and is
the familiar solution found by perturbation expansion around $g=0$.
The perturbative solution vanishes at $J = 0$ is regular in $g$ at
$g=0$. The other solutions break reflection symmetry and are singular at
$g=0$. The symmetry breaking solutions always show singularities as the coupling
vanishes. You often have to be very careful to not miss these.

Now returning to higher dimensions (which, in principle, can be
constructed from zero dimensional solutions) a new element occurs -
namely the Nambu Goldstone Theorem. This (roughly) states that in more
than 2 spacetime dimensions) that if a \emph{charge} associated with a
\emph{conserved current} in a relativistic field theory does not
\emph{destroy} the vacuum $\Rightarrow$ the theory has zero mass
excitations. At first glance the theorem seems to be true using exact
results of QFT without use of \emph{any} approximation
techniques. Furthermore, in an interacting theory, you can not break a
symmetry using a coupling constant based perturbative expansion around
zero coupling strength.

\textbf{What is Goldstone's Theorem good for?}
As mentioned above, before 1964 it appeared to be an impediment to creatively use symmetry breaking
because of the dearth of physical massless particles.

This is where I come in: After Bjorken gave a talk (1962) at Harvard,
my thesis advisor, Walter Gilbert (Nobel Laureate Chemistry 1980),
suggested that I look at Bjorken's proposed model of E\&M \cite{jb;1963} --- a
variant of the Nambu--Jona-Lasinio model with interaction

\begin{equation*}
  g\, (\bar\psi\, \gamma^{\mu}\, \psi)\, (\bar\psi\, \gamma_{\mu}\, \psi) \; .
\end{equation*}
The current is required to have non-vanishing vacuum expectation.
The symmetry that is broken is Lorentz symmetry --- relativistic
invariance.

In my Ph.D thesis, I showed that BJ's basic conclusion that this
theory is equivalent to QED is correct. Careful calculation shows that
the Lorentz symmetry breaking is trivial and does not manifest itself
in a physically observable way \cite{ggff1;1964,ggff2;1964}. This is a surprise since in
\emph{coupling constant perturbation theory} this interaction leads
to hopelessly divergent results. In fact, this calculation provides results in a different [quantum]
phase corresponding to symmetry breaking boundary conditions and is an
entirely different solution than the non-existent coupling constant
perturbation expansion. This is the direct analog to the multiple solutions of  ordinary
differential equations previously discussed.

Despite the fact that by this time Schwinger \cite{sch1;1962} had
argued that there was no dynamical reason for the photon to have zero
mass, I thought from what I had learned about the Bjorken model that I
could construct a symmetry breaking argument that disproved Schwinger's
conclusions and demonstrated that conventional E\&M must, in general,
have massles photons. I included this in my thesis. My argument was
wrong and, fortunately, Coleman detected this in my (1963) thesis
presentation. I removed the offending chapter in the final version.

Somewhat before my thesis was finished I had discussed a related
project with Gilbert. He made the observation that the action of a
massless \emph{scalar} particle $(B)$ and a massless \emph{vector}
particle $(A^{\lambda})$ with the simple ``interaction''

\begin{equation*}
  g\, A^{\lambda}\, \left(\pa_{\lambda}B - g\,A_{\lambda}\right)
\end{equation*}
produces a free spin 1 field with mass $g^2$. This can be
\emph{anticipated} by \emph{counting} degrees of freedom and noting
that $g$ carries the dimension of mass (this model has a conserved
current and a trace of gauge invariance).

I told David Boulware about this and he spoke to Gilbert and they
wrote a paper including some additional discussion about massless limits of massive vector
meson theories \cite{bg;1962}. Thus, at this time,
the 2-dimensional Schwinger model (E\&M in 2 dimensions) showed that
gauge theories need not have zero mass and the BG model in 4
dimensions \emph{confirmed} this again. It is a \emph{easy} step from
the BG model to the lowest approximation used in the GHK paper. They
are essentially the same! With hindsight, all the ingredients for the
GHK paper were available at Harvard in 1962!

During my time at Harvard, I was talking with Dick Hagen an
undergraduate friend at MIT and then a Physics graduate student at MIT
and already my co-author on our first physics research paper \cite{gh1;1963}.

\begin{center}
  \includegraphics[scale=0.27]{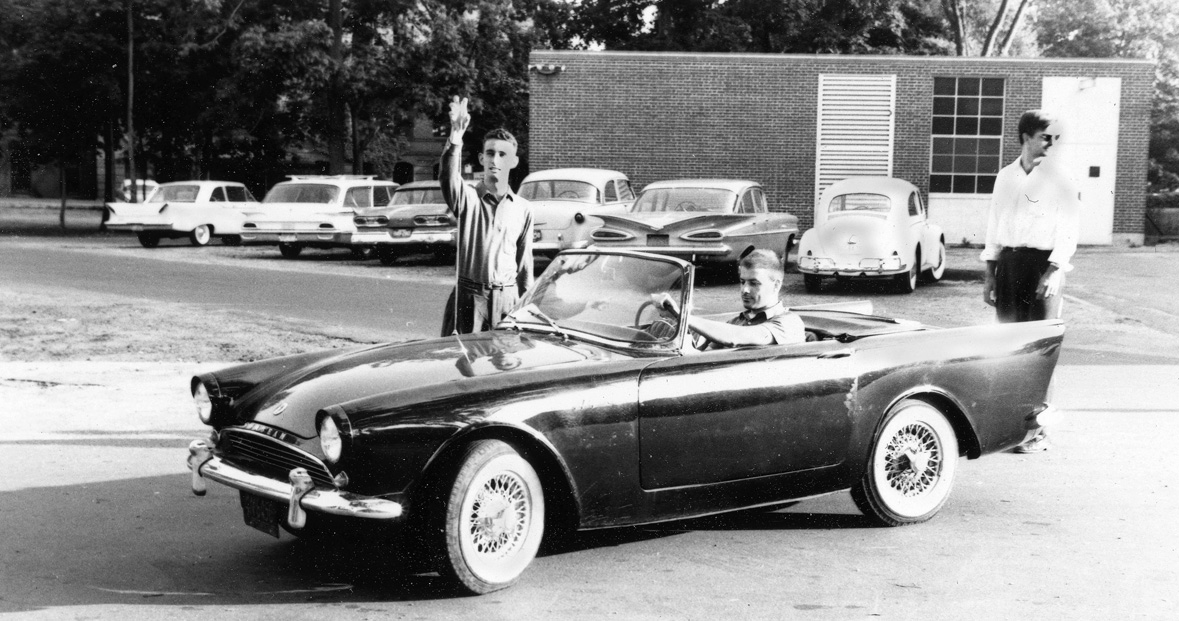}

  A Working Afternoon: Young Baron May of Oxford, Guralnik and Hagen 1961
\end{center}

In 1963 Hagen took a postdoctoral position at the University of
Rochester (and is still there). We continued collaborating. He became
interested in complicated expensive and unreliable but beautiful
machinery - see the photos below. Also he became an expert on how to minimize living costs. For a while, I thought he might be
thinking of becoming a very ``hands on'' experimentalist.

\begin{flushleft}
  \includegraphics[scale=25.1]{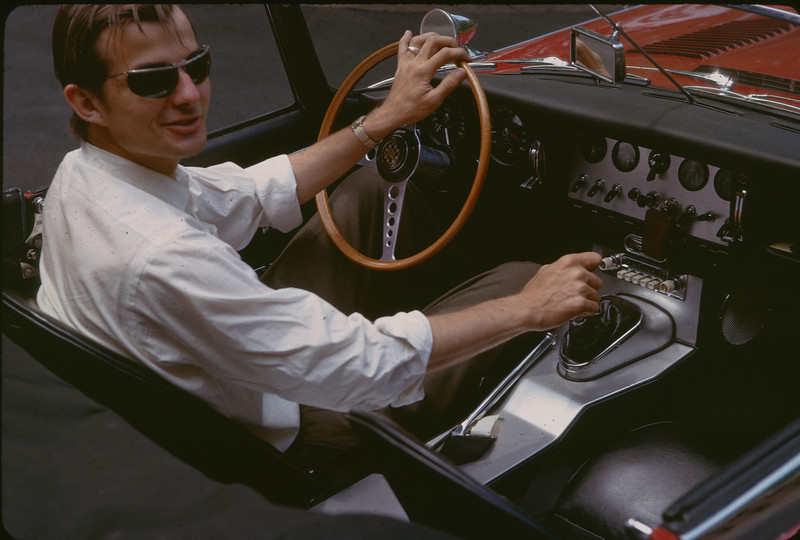}
\end{flushleft}

\begin{center}
  \includegraphics[scale=25.1]{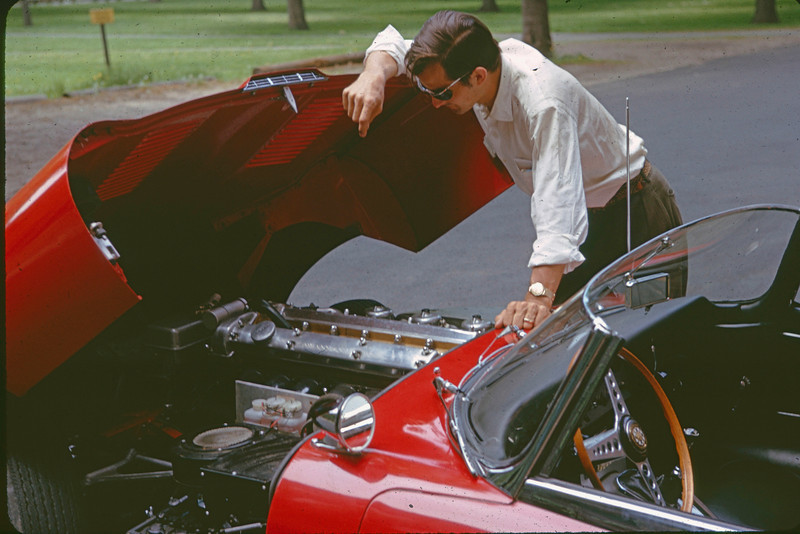}
\end{center}

\begin{flushright}
  \includegraphics[scale=25.1]{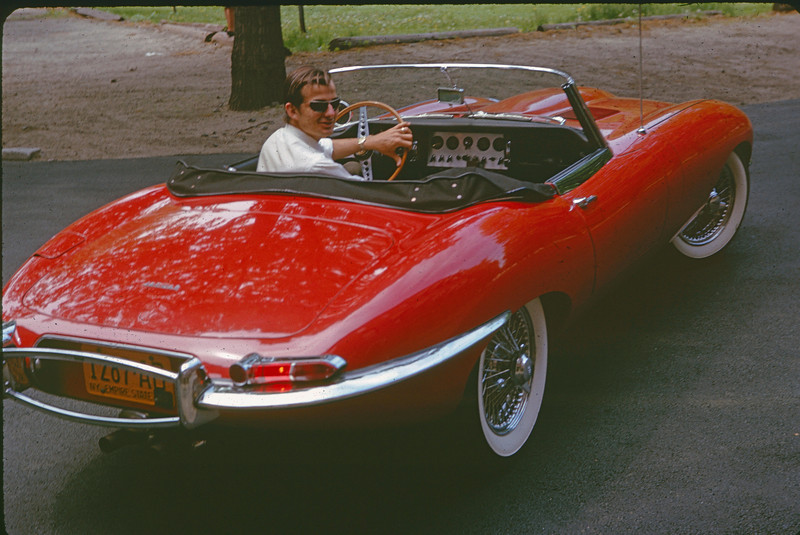}
\end{flushright}

I went to Imperial College (after being rejected by CERN) at the
beginning of 1964 with a new NSF postdoctoral fellowship and the
certainty that \emph{something} \emph{interesting} happened with
gauge theories and \emph{symmetry} \emph{breaking}. IC was probably
the best High Energy Theory place in the world at that time and I met
a fantastic bunch of physicists there. The ones I interacted with the
most were Tom Kibble, Ray, Streater, John Charap, and to a lesser
degree Paul Matthews and Abdus Salam.

\begin{center}
  \includegraphics[scale=0.6]{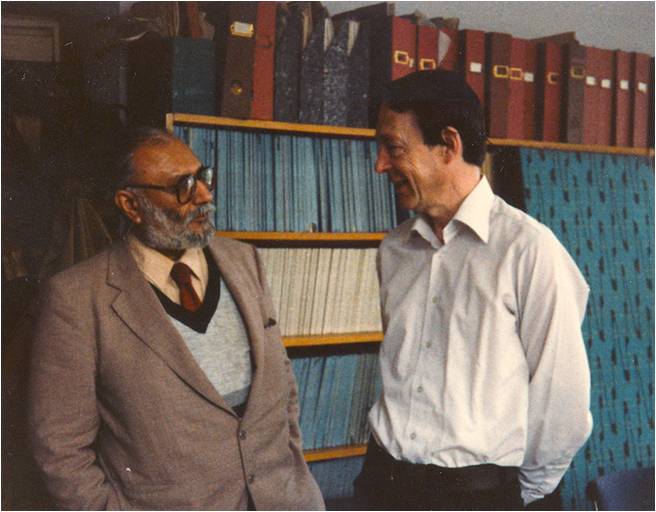}

  Salam and Kibble -like the Harvard and MIT crowd, the IC people were very serious
\end{center}

I quickly learned that while Harvard was relatively safe field theory
ground, protected by Schwinger's large (but indifferent) umbrella, the
idea that there was even such a thing as symmetry breaking in field
theory was not universally accepted --- even at IC where Salam
with Goldstone and Weinberg had already published their nice paper on
these ideas \cite{gsw;1962}.

Ray Streater (an axiomatic or constructive field theorist) stated that
his community did not believe that symmetry breaking was possible. A
lot of arguing and careful construction of a free model
\emph{convinced} him that the axioms were too restrictive. Later. he
published a paper on this \cite{rs;1965}, which amusingly got a lot more attention
than the paper I published in PRL giving the simple free example and a
significant example of the Goldstone theorem in gauge theory but  with
an incomplete analysis of the resulting gauge structure \cite{ggfu;1964}.

The understanding of my oversight in this paper (which, incidentally,
was also caught by Dave Boulware) was the final key to the GHK
understanding, within the context of the Goldstone theorem and without
resorting to perturbation theory, why symmetry breaking in a gauge
theory, does not require massless particles. In gauge theories the
assumptions of the Goldstone theorem are easily (and always) violated in physical
gauges, or equivalently, are only applicable to non-physical
excitations.

I begin by re-examining my earlier PRL paper. As I proceed, it will be
clear why my explanation is the basis of the GHK paper. The proof in
QED is straightforward: There is an asymmetric conserved tensor
current,
\begin{align*}
  J^{\mu\, \nu} &= F^{\mu\, \nu} - x^{\nu}\, J^{\mu} \\
  \pa_{\mu}\, J^{\mu\, \nu} &= 0 \\
  \Rightarrow\; Q^{\nu} &= \int d^3x \left[ F^{0\, \nu} - x^{\nu}\, J^0 \right] \\
  \text{and } \frac{d}{dt}\, Q^{\nu} &= 0 \; .
\end{align*}
We use the gauge $\vec\nabla\cdot\vec{A} = 0$ (a very natural gauge
in operator QED) so that we only deal with physical excitations.

By the commutation relations it is easily seen that this requires
\begin{equation*}
  \ipop{0}{\comm{Q^{k}}{A^{l}(\vec{x}, t)}}{0} = (\text{non-zero constant}) \; .
\end{equation*}
However, direct calculation using spectral representations show that
this expression is time-dependent for $e \neq 0$!

\textbf{What went wrong?} The radiation gauge is not explicitly
Lorentz invariant, and we cannot use \emph{causality} arguments to
prove that the \emph{commutator} above is \emph{confined} to a local
region of space-time. This means that, even though $\pa_{0} J^{0\, 0} + \pa_{k} \, J^{0\, k}
= 0$, we cannot  neglect surface integrals of $J^{0\, k}$. It follows
that our weird  \textbf{charge  leaks out of any volume!} This leads us, at
once, to re-consider the proof of Goldstone's theorem.

\textbf{What have we learned?}

Goldstone's theorem is true for a manifestly covariant theory,
i.e., a theory where $\pa_{\mu} J^{\mu} = 0$ and \emph{surface terms}
vanish \emph{fast enough} so that
\begin{equation*}
  \begin{split}
    & \ipop{0}{\comm{\int d^3x (\pa_{\mu}\, J^{\mu})}{\text{(local operator)}}}{0} = \\
    &\quad = \frac{d}{dt}\ipop{0}{\comm{\int d^3x\, J^0}{\text{(local operator)}}}{0} \; .
  \end{split}
\end{equation*}

That is to say:
\begin{equation*}
  Q = \int d^3x\, J^0
\end{equation*}
has a zero mass particle in its spectrum. This includes
electromagnetism with the special charge introduced above if you
re-gauge to a manifestly covariant gauge. However, in this case, you
can demonstrate exactly that the zero mass particles are gauge
excitations. Note that these are very general statements: Goldstone's
theorem need not require physical zero mass states in any gauge theory
(and it does not). This is because these theories are made to be
relativistic by introducing extra gauge degrees of freedom. Indeed,
the Goldstone bosons are always nonphysical.

There is no reason for the photon to be massless in normal QED, but
the smallness of the coupling constant and hence the applicability of
perturbation theory. We can see an approximate example of the
failure of Goldstone's theorem by looking at the action
\begin{align*}
  L &= -\frac{1}{2}\,
  F^{\mu\, \nu}\, (\pa_{\mu}A_{\nu} - \pa_{\nu}A_{\mu}) +
    \frac{1}{4}\, F^{\mu\, \nu}\,F_{\mu\, \nu} + \phi^{\mu}\pa_{\mu}\phi  + \\
  &\quad+ \frac{1}{2}\, \phi^{\mu}\phi_{\mu} + i\,e_0\,
    \phi^{\mu}\, \mathbf{q}\, \phi\, A_{\mu} \\
  \mathbf{q} &= \sigma_2 \\
  \phi &= (\phi_1, \phi_2) \\
    \phi_{\mu} &= (\phi^{\mu}_{1},\phi^{\mu}_{2})
\end{align*}

This is the Lagrangian for scalar electrodynamics. It is a very
non-trivial interacting theory characterized by a conserved
current. It is renormalizable in the coupling constant expansion with
an induced $\phi^{4}$ interaction. No other non-trivial $\phi^{n}$
interaction can be added to it and keep it renormalizable. Here and in our paper we use the
convention followed at Harvard and MIT by the ``Schwinger School'' and do not introduce (unneeded) explicit
counter-terms. This property and much of the general behavior of scalar electrodynamics was examined by Salam in 1951
\cite{as;1951}. We want to look for solutions other than the coupling constant expansion.
At the time, it was very natural for us to put in a source for $\phi$
and order an iterative expansion by the number of derivatives with
respect to the source. A variant of this method was used in my thesis to study the Bjorken model.
However, we chose in GHK to keep things simple by writing down the equivalent leading order expansion
in operator form. Note our approach is fully quantum mechanical and that the Goldstone theorem arguments
depending on commutation relations are quantum mechanical as well. Here, the concept of a particle only has
meaning in quantum mechanics and in no way is it necessary to calculate higher order corrections to have
quantum results as has been argued elsewhere.

The leading approximation is obtained by replacing $i\, e_0\,
\phi^{\mu}\, \mathbf{q}\, \phi\, A_{\mu}$ in the Lagrangian by
$\phi^{\mu}\, \eta\, A_{\mu}$. (The result is essentially the
Boulware--Gilbert action with an extra scalar field). This ``reduced
Lagrangian'' results in the linearized field equations:
\begin{align*}
  F^{\mu\, \nu} &= \partial^{\mu} A^{\nu} - \partial^{\nu} A^{\mu} \; ;\\
  \partial_{\nu} F^{\mu\, \nu} &= \phi^{\mu}\, \eta\; ; \\
  \phi^{\mu} &= -\partial^{\mu}\phi - \eta\, A^{\mu} \; ; \\
  \partial_{\mu}\phi^{\mu} &= 0\; .
\end{align*}
These equations are soluble, since they are (rotated) free field
equations. The diagonalized equations for the physical degrees of freedom
are:
\begin{align*}
  (-\partial^2 + \eta_1^2)\, \phi_1 &= 0 \; ;\\
  -\partial^2 \phi_2 &= 0 \; ;\\
  (-\partial^2 + \eta_1^2)\, A_k^T &= 0 \; .
\end{align*}

For convenience, we have made the assumption that $\eta_1$ carries the
full value of the vacuum expectation of the scalar field (proportional
to the expectation value of $\phi_2$).  The superscript ${}^T$ denotes
the transverse part.  The two components of $A_k^T$ and the one
component of $\phi_1$ form the three physical components of a massive
spin-one field while $\phi_2$ is a spin-zero field. As previously
mentioned, the Goldstone theorem is not valid, so there is no
resulting massless particle associated with this theorem. If the
Goldstone theorem were valid, $\phi_1$ would be massless. It is very
important to realize that it is an artifact of the lowest order
approximation for the above action that $\phi_2$ is massless. The
excitation spectrum of this field is not constrained by any
theorem. It is obvious to anyone who has calculated in QFT that there
are higher order corrections and that, in fact, they diverge in 4
space-time dimensions and to get a final result that this theory need
mass and coupling constant renormalization using experimental
input. If this were not so, the theory would be basically inconsistent
and free. This was obvious in 1964 and it is now. While knowledge
beyond that which existed in 1964 is not necessary to make the
preceding statements, if the reader wants to understand this in more
detail than these obvious statements without calculating themselves I
suggest that you refer to the famous paper by Coleman and E. Weinberg
\cite{cw;1972}

At this stage, it might be thought that we have written down an
interesting, but possibly totally uncontrolled, approximation. There
is no \emph{a priori} reason to believe that this is even a meaningful
approximation. The main result, that the massless spin-one field and
the scalar field unite to form a spin-one massive excitation, could be
negated by the next iteration of this approximation. However, this
approximation meets an absolutely essential criterion that makes this
unlikely. While the symmetry breaking removes full gauge invariance,
current-conservation, which is the fundamental condition, is still
respected. This is clear from the above linearized equations of
motion. We can directly demonstrate that the mechanism, described
earlier in this note for the failure of the Goldstone theorem, applies
in this approximation.

The internal consistency and the consistency with exact results gives
this approximation credence as a leading order of an actual
solution. It is, in fact, not hard to make this the leading order of a
well defined approximation scheme.

This solution of the action describes a 3-degree of freedom
spin 1 particle and a 1-degree of freedom spin 0 particle.
Goldstone's theorem does not apply, even though the current
\begin{equation*}
  J^{\mu} = i\, e_0\, \phi^{\mu}\, \mathbf{q}\, \phi = \phi^{\mu}\cdot \eta
\end{equation*}
is conserved.

%%%%%%%%%%%%%%%%%%%%%%%%%%%%%%%%%%%%%%%%%%%%%%%%%%%%%%%%%%%%%%%%%%%%%%%%%%%%%%%
%%%%%%%%%%%%%%%%%%%%%%%%%%%%%%%%%%%%%%%%%%%%%%%%%%%%%%%%%%%%%%%%%%%%%%%%%%%%%%%
%%%%%%%%%%%%%%%%%%%%%%%%%%%%%%%%%%%%%%%%%%%%%%%%%%%%%%%%%%%%%%%%%%%%%%%%%%%%%%%

\section{Summary and Comparisons}

The GHK paper addresses two major issues in detail.  As a general
issue it explains in detail why gauge theories do not
intrinsically require zero mass particles. This emphatically does not
depend on a specific model, but is a consequence of the ``leakage'' of
appropriate charges out of any surface. This is a fully quantum
mechanical proof and it is exact, ultimately depending only on the
general structural form of the radiation gauge propagator.  As a
specific issue the GHK paper examine a specific theory - Scalar
Electrodynamics with the charge symmetry broken by requiring non-
vanishing expectation value of the scalar field. We demonstrate that
in a (non-coupling constant perturbative) self consistent leading
order approximation that this theory has no Goldstone boson and that
the original degrees of freedom combine to produce a massive vector
particle with the mass depending on the symmetry breaking parameter
and a single component real scalar particle with no intrinsic
constraint on its mass.

There has been considerable recent discussion about the relative merit
and content of the work of the three groups that first analyzed
spontaneous symmetry breaking as now used in the unified theory of
weak and electromagnetic interactions. Some of this discussion has
been quite misleading and even wrong. The fact that the papers are
over 45 years old means that for fair evaluation, it is important to
be aware of both the sophistication and the limits of our
understanding at that time. With this in mind, I will do my best to
make clear some of the important differences between the three sets of
work. An essential point to note is that understanding the
possibility of ``avoiding the Goldstone Theorem'' (the motivation for
all of these papers) encompasses two very distinct parts. The first
part is to find a mechanism that makes it possible to spontaneously
break symmetry in relativistic theories without producing massless
particles.  The second part is to find a specific example. It is
possible to do either of these parts independently without an
understanding of the other.

In regards to the mechanism, Brout and Englert make some comments, but
give no quantitative argument. In the Higgs PRL paper, there is only
the specific scalar electrodynamics example. In his physics letters
paper Higgs observed that an argument given by Gilbert to negate the
Goldstone theorem includes a term with a fixed vector such as can be
found in radiation gauge electrodynamics. He makes no follow-up or
extension of this argument in his PRL paper.

The GHK paper does an extensive analysis of the mechanism and shows
that in broken gauge theories in the radiation gauge that charge leaks
out of any volume and consequently is not conserved since currents
continue to exist in the limit of volumes tending toward infinity. This negates
the basic assumption of a conserved charge that is required by the
Goldstone theorem. Since the radiation gauge removes all gauge modes,
the Goldstone theorem, which applies in manifestly Lorentz invariant gauges, can only require massless
gauge particles. This is discussed in considerable detail in our
paper.  This is an exact model independent result not limited to the
explicit scalar field interaction and which, in particular, applies to
composite models. When combined with the fully analyzed results of my
previous related PRL paper it shows that there is no requirement that
the photon have zero mass except when coupling constant perturbation
theory is valid (this was probably first conjectured by Schwinger). A
method for examining the photon mass through the Goldstone Theorem was
introduced in my early PRL paper, used, but not discussed, to build the
later GHK paper and analyzed in detail in my ``Feldafing'' talk given
in the summer of 1965 \cite{hc;1965}. I believe that it is correct to say that GHK
was the only group to actually analyze the general mechanism. This observation is likely
to be particularly important if the LHC does not reveal a fundamental scalar boson.

All three groups used the scalar electrodynamic Lagrangian to provide
an example (in a first approximation) of a gauge theory with a
broken symmetry that yields a massive spin one particle but their
analysis is different.

EB does not fully construct the lowest order approximation. They do
not show the ``Higgs Boson'' and while they do argue that their
approximation respects current conservation it seems to me that their
analysis is in general quite incomplete. They do make the essential
identification of the massive parameter. They further ``assume that
the application of the theorem of Goldstone, Salam and Weinberg is
straightforward and thus that the propagator of the field $\phi_2$ which
is ``orthogonal'' to $\phi_1$ has a pole at q=0 which is not isolated.''
It is true that in a covariant gauge that the theorem is valid and
that there is a zero mass pole. That pole is not physical and is
purely gauge, hence unphysical. It does not seem to me that
is what EB are saying but I leave the interpretation to the reader.

The Higgs PRL is probably more complete than the EB paper
although entirely classical. It  fails to write down the full
solution to the leading order equations displayed. He does have (the classical analogue) of a zero mass
solution which is not mentioned or displayed. This solution is pure gauge but it needs to be noticed and
properly handled. Higgs includes an explicit scalar
self coupling in his model and writes down the equations for both scalar
degrees of freedom. As mentioned above, GHK fully describes why the Goldstone boson is a gauge only
excitation in manifestly covariant formulations and is not present in
the physical radiation gauge. The results are exact and independent of the specific model
analyzed in an approximation.

Recently it has been claimed that the GHK paper
does not have the ``Higgs boson''. This claim astonishes us. We, far
more than any of the other groups, keep very careful track of the
degrees of freedom of our scalar electrodynamics model. On the bottom
of the right column of page 586 of the GHK paper are the three
equations for the leading order approximations to the 4 physical
degrees of freedom. We observe that the two degrees of freedom of the
vector field combine with one scalar boson to form the three degrees
of freedom of a massive spin one vector field. There is one remaining
scalar field, $\phi_2$ in our notation, which in our approximation has
zero mass. That this mass is zero has absolutely nothing to do with any
dynamical constraint including the Goldstone theorem. The Goldstone
theorem, if valid here, would only constrain the mass of $\phi_1$. The
zero mass is an artifact of how we pick the explicit action and the
leading order approximation. This is different from the Higgs paper in
that he puts in an explicit pure scalar interaction.  In a 4
dimensional renormalizable theory that interaction is limited to being
pure quartic. As was our practice mirroring that commonly used by
Schwinger and associates, we did not put in this explicit quartic term
in scalar electrodynamics but were fully aware that such a term is
generated in higher approximations. Ultimately because, of renormalization, the GHK
choice of the action is operationally identical to the one used by Higgs.

 In summary, our purpose was to show that the Goldstone
theorem did not constrain physical mass in scalar gauge theories. We
demonstrated this generally and in a specific example. The mass of
$\phi_2$ happens to be zero in leading order, but as was obvious to us
and every other experienced field theorist of that time, this would
change order by order as the theory was iterated in a manner closely
related to how it changes in unbroken scalar
electromagnetism.

\begin{center}
  \includegraphics[scale=0.26]{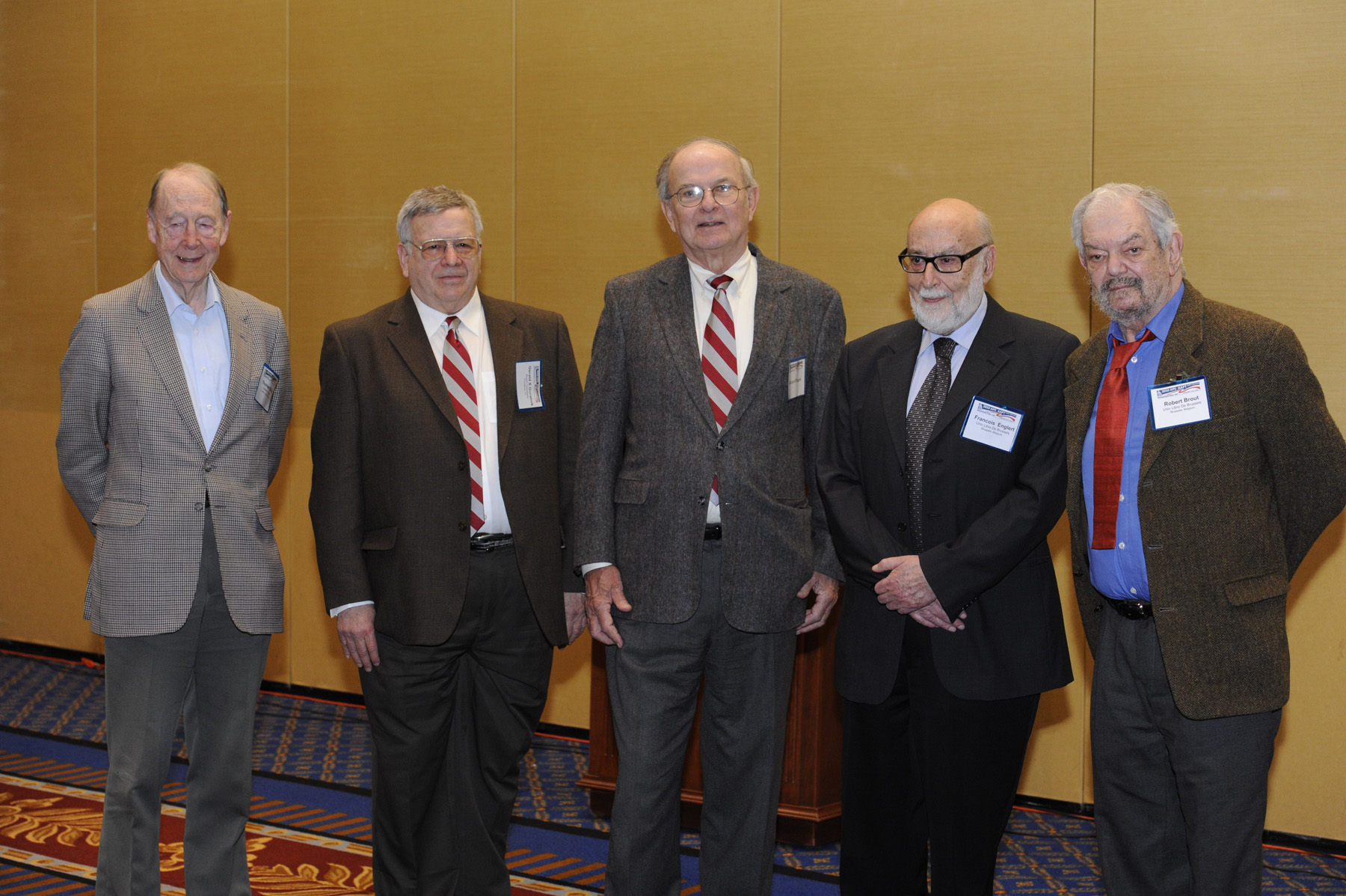}

  ``Gang of five'' Sakurai Prize 2010
\end{center}

The following show Tom Kibble and Gerald Guralnik hoping to obtain wisdom by emulation:

\begin{flushleft}
  \includegraphics[scale=0.10]{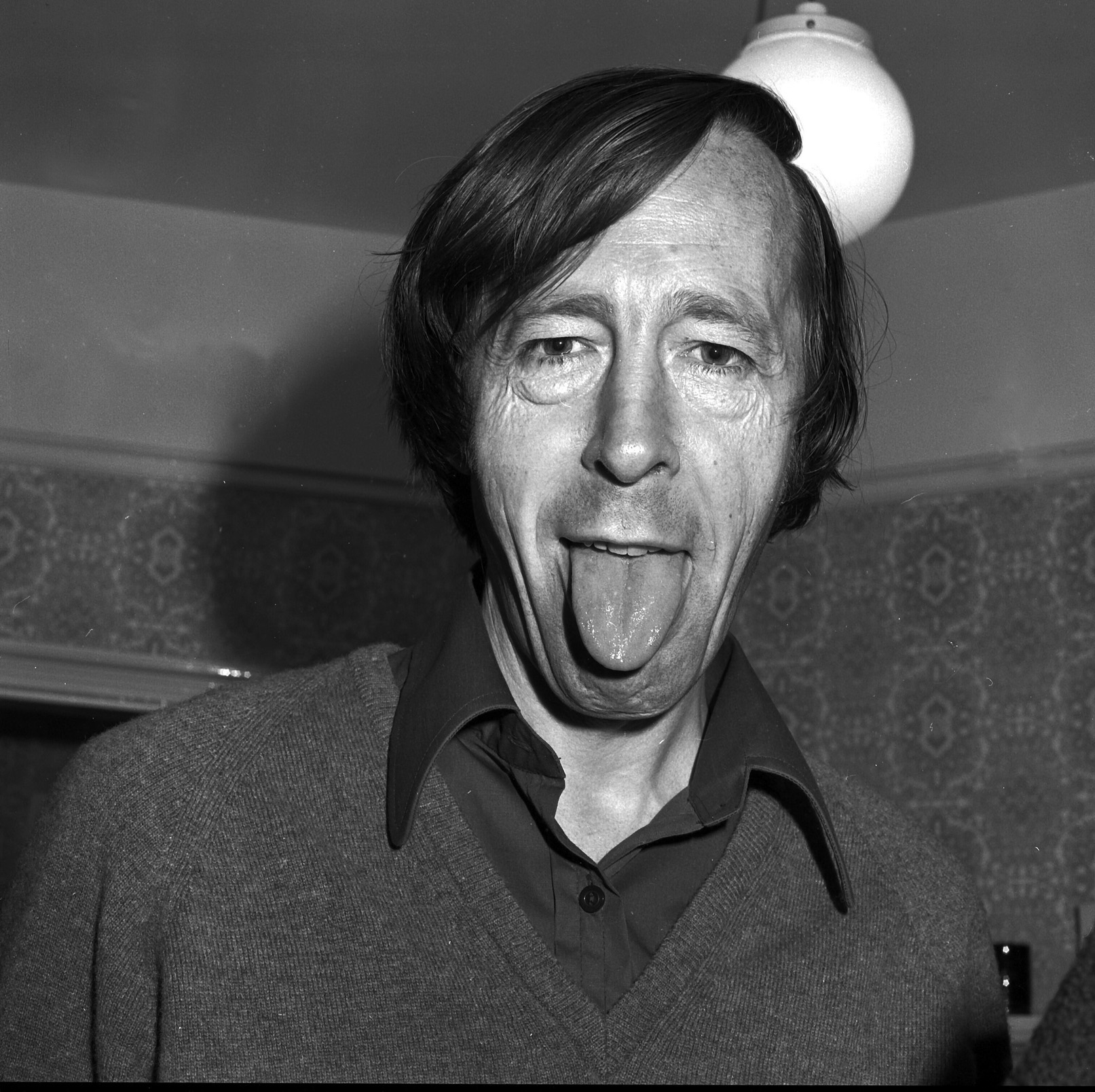}
\end{flushleft}

\begin{flushright}
  \includegraphics[scale=0.16]{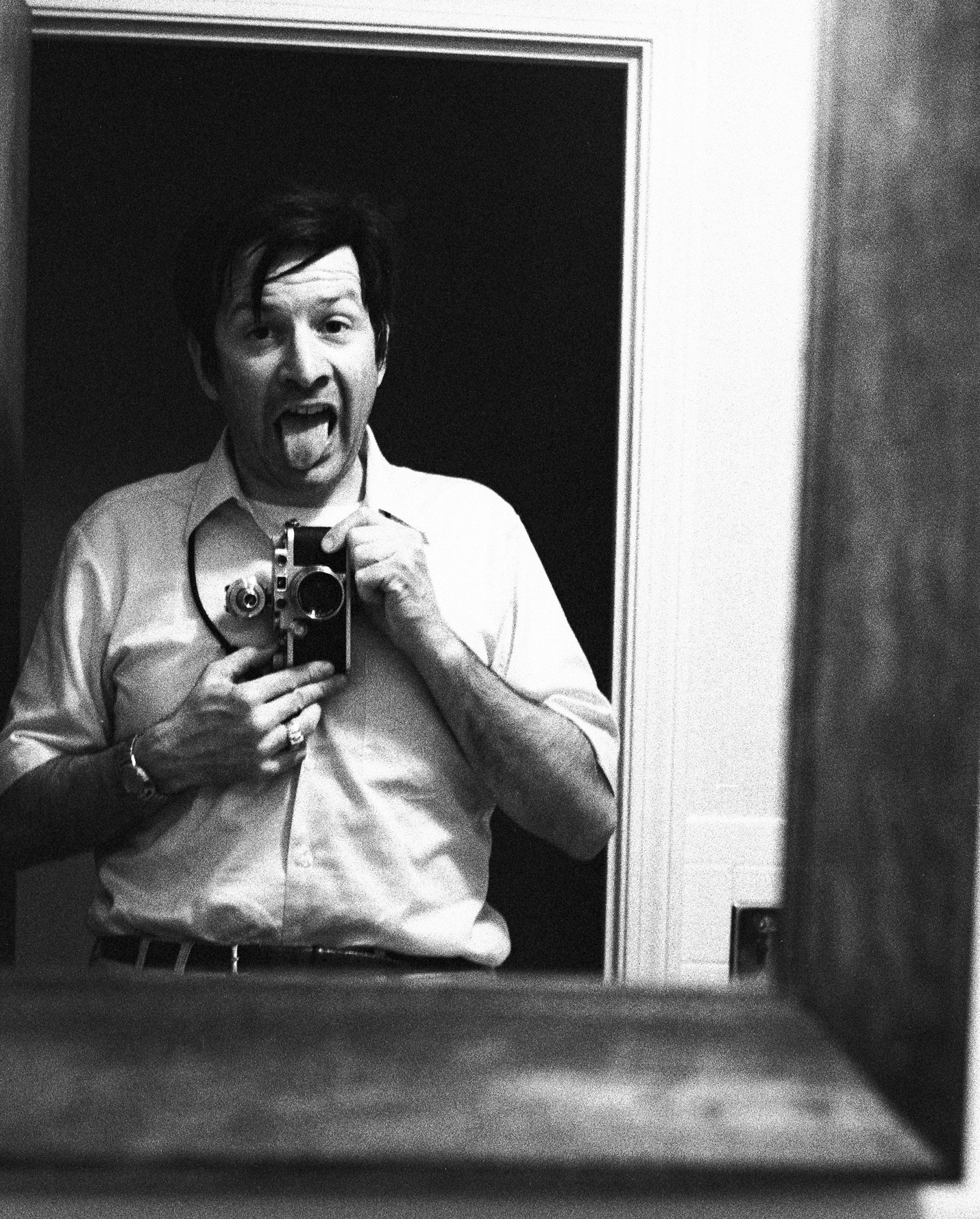}
\end{flushright}
\begin{acknowledgments}
I wish to thank Richard Hagen and Tom Kibble for the collaboration that made
the GHK paper possible, as well as for discussions over the years that have
taught me so much. This work is supported in part by funds provided by the US Department
of Energy (\textsf{DoE}) under \textsf{DE-FG02-91ER40688-TaskD}.
\end{acknowledgments}

\bigskip % extra skip inserted

%%
%%%

\begin{thebibliography}{99}
\bibitem{gg;2009} G.S. Guralnik, ``The History of the Guralnik, Hagen and
  Kibble Development of the Theory of Spontaneous Symmetry Breaking and Gauge
  Particles'', IJMPA. {\bf 24}, 2601 (2009)

\bibitem{ggg;1996} Santiago Garcia, Zachary Guralnik and G.S. Guralnik, ``Theta
  Vacua and Boundary Conditions of the Schwinger Dyson Equations'' e-print:
  hep-th/9612079.
\bibitem{ggzg;2007} G. Guralnik and Z. Guralnik ``Complexified path integrals
  and the phases of quantum field theory.'' e-Print: arXiv:0710.1256 [hep-th].
\bibitem{ghk;1964} G.S. Guralnik, C.R. Hagen and T.W.B. Kibble, ``Global
  Conservation Laws and Massless Particles'',  Phys. Rev. Lett. {\bf 13}, 585 (1964).
\bibitem{eb;1964} F. Englert and R. Brout, ``Broken Symmetry and the Mass of
  Gauge Vector Mesons'', Phys. Rev. Lett. {\bf 13}, 321 (1964).
\bibitem{phpl;1964} Peter Higgs (1964), ``Broken Symmetries, Massless Particles
  and Gauge Fields'', Physics Letters {\bf 12}, 132.
\bibitem{ph;1964} Peter W. Higgs, ``Broken Symmetries and the Masses of Gauge
  Bosons''. Phys. Rev. Lett. {\bf 13}, 508 (1964).
\bibitem{sw;1967} Steven Weinberg, ``A Model of Leptons'', Phys. Rev. Lett.{\bf 19}, 1264
  (1967).
\bibitem{as;1967} Abdus Salam, Unpublished Imperial College Lecture
\bibitem{gh1;1963}  G.S. Guralnik and C.R. Hagen, ``Regge Poles in Relativistic
  Wave Equations'', Phys. Rev. {\bf 130}, 1259 (1963).
\bibitem{ggff1;1964} G.S Guralnik, ``Photon as a Symmetry Breaking Solution to
  Field Theory I'', Phys. Rev. {\bf 136}, 1404 (1964).
\bibitem{ggff2;1964} G.S. Guralnik, ``Photon as a Symmetry Breaking Solution to
  Field Theory II'', Phys. Rev. {\bf 136}, 1417 (1964).
\bibitem{ym;1954} Yang, C. N.; Mills, R. (1954), "Conservation of Isotopic Spin and
 Isotopic Gauge Invariance". Physical Review {bf 96 }(1): 191.
\bibitem{ss;1955} Shaw, Ronald. Cabridge Ph.D thesis 1955.
\bibitem{sak;1960} J.J. Sakurai, Ann. Phys. 11, 1 (1960).

\bibitem{mg;1962} Murray Gell-Mann ``Symmetries of Bayons and Mesons'', Phys. Rev. {\bf 125}, 1067 (1962)
\bibitem{yn;1961} Y. Ne'eman, ``Derivation of Strong Interactions from
  a Gauge Invariance'', Nuclear Phys., {\bf 26}, 222 (1961).
\bibitem{gm;1964} M. Gell-Mann ``A Schematic Model of Baryons and Mesons''. Physics Letters.{\bf 8}, 214 (1964)
\bibitem{gz;1964} G. Zweig ``An SU(3) model for strong interaction
  symmetry and its breaking'' CERN preprint (1964).
\bibitem{wag;1964} O. W. Greenberg, ``Spin and Unitary-Spin Independence in a
  Paraquark Model of Baryons and Mesons'', Phys. Rev. Lett. {\bf 13}, 598 (1964).
\bibitem{gmf;1958} R.P. Feynman and M. Gell-Mann ``Theory of the Fermi Interaction'',
Phys. Rev. {\bf 109}, 193 (1958)
%\bibitem{nl;1962} Y.Nambu and D. Lurie, ``Chirality Conservation and Soft Pion
%  Production'', Phys. Rev. {\bf 125}, 1429 (1962).

\bibitem{nb;1960} Y. Nambu and G. Jona-Lasinio, ``Dynamical Model of
  Elementary Particles Based on an Analogy with Superconductivity. I'',
  Phys. Rev. {\bf 122}, 345 (1961).

\bibitem{nb;1961} Y. Nambu and G. Jona-Lasinio, ``Dynamical Model of
  Elementary Particles Based on an Analogy with Superconductivity. II'',
  Phys. Rev. {\bf 124}, 246 (1961).

\bibitem{jg;1961} J. Goldstone, Nuovo Cimento {\bf 19}, 154 (1961).
\bibitem{gsw;1962} J. Goldstone, A Salam and S Weinberg, ``Broken
  Symmetries''. Physical Review {\bf 127}, 965 (1962).
\bibitem{dyson;1952} F.J. Dyson, ``Divergence of Perturbation Theory in Quantum
  Electrodynamics'', Phys. Rev. {\bf 85}, 631 (1952).
\bibitem{ggfu;1964} G.S. Guralnik, ``Naturally Occurring Zero Mass Particles and Broken
Symmetries'',  Phys. Rev. Lett. {\bf 13}, 295 (1964).
\bibitem{jb;1963} J. Bjorken, ``A dynamical origin for the electromagnetic field'',
  Annals Phys. {\bf 24}, 174 (1963).
\bibitem{sch1;1962} Julian Schwinger, ``Gauge Invariance and Mass'',
  Phys. Rev. {\bf 125}, 397 (1962). (Volume 125, Issue 1.)
\bibitem{bg;1962} David G. Boulware, Walter Gilbert, ``Connection between Gauge
  Invariance and Mass'', Phys. Rev. {\bf 126}, 1563 (1962).
\bibitem{sch2;1962} Julian Schwinger, ``Gauge Invariance and Mass. II'',
  Phys. Rev. {\bf 128}, 2425 (1962).
\bibitem{rs;1965} R.F. Streater, ``Spontaneously broken symmetry in axiomatic
  theory'', Proc. Roy. Soc. Lond. {\bf A287},510 (1965).

\bibitem{twbk;1967} T.W.B. Kibble, ``Symmetry breaking in nonAbelian gauge theories''
Phys.Rev.{\bf155},1554 (1967).

\bibitem{ghk;1968} G.S. Guralnik, C.R. Hagen, and T.W.B. Kibble, ``Symmetry
  Breaking and the Goldstone Theorem".
  {Advances in Particle Physics - Vol. II}, p. 567; Interscience
  Publishers Inc., New York (1968).

\bibitem{pwa;1963} P. W. Anderson, ``Plasmons, Gauge Invariance, and Mass'',
  Physical Review {\bf 130}, 439 (1963).
\bibitem{wg;1964} W. Gilbert, ``Broken Symmetries and Massless Particles'',
  Physical Review Letters {\bf 12}, 713 (1964).
\bibitem{rvl;1965} R. V. Lange ``Goldstone Theorem in Nonrelativistic
  Theories'', Phys. Rev. Lett. {\bf 14}, 3 (1965).
\bibitem{phr;1965} Peter Higgs, ``Spontaneous Symmetry Breakdown without
  Massless Bosons'', Physical Review {\bf  145},1156 (1966).
\bibitem{as;1951} Abdus Salam, ``Renormalized S-Matrix for Scalar Electrodynamics'' Phys. Rev. {bf 86}, 731 (1951)

\bibitem{cw;1972} Sidney Coleman and Erick Weinberg ``Radiative
  Corrections as the Origin of Spontaneous Symmetry Breaking'',
  Physical Review. {\bf 7}, 1888 (1972)

\bibitem{hc;1965} G.S. Guralnik, ``Gauge Invariance and the Goldstone Theorem'',
  Proceedings of Seminar on Unified Theories of Elementary Particles, p. 91,
  Munich (1965).




\end{thebibliography}
\end{document}